# Quasi-continuous transition from a Fermi liquid to a spin liquid

**Authors:**

Tetsuya Furukawa[1*†], Kazuhiko Kobashi[1], Yosuke Kurosaki[1], Kazuya Miyagawa[1], Kazushi Kanoda[1*]

**Affiliations:**

[1]Department of Applied Physics, University of Tokyo, 7-3-1 Hongo, Bunkyo-ku, Tokyo 113-8656, Japan.

[*]Corresponding author: e-mail: tetsuya.furukawa@rs.tus.ac.jp (T.F), kanoda@ap.t.u-tokyo.ac.jp (K.Kanoda)

[†]Current address: Department of Applied Physics, Tokyo University of Science, Niijyuku 6-3-1, Katsushika-ku, Tokyo 125-8585, Japan

**The Mott metal-insulator transition—a drastic manifestations of Coulomb interactions among electrons—is the first-order transition of clear discontinuity, as shown by various experiments[1] and the celebrated dynamical mean-field theory[2]. Recent theoretical works, however, suggest that the transition is continuous if the Mott insulator carries an exotic spin liquid with a spinon Fermi surface[3,4]. Here, we demonstrate the case of a quasi-continuous Mott transition from a Fermi liquid to a spin liquid in an organic triangular-lattice system $\kappa$-(ET)$_2$Cu$_2$(CN)$_3$. Transport experiments performed under fine pressure tuning find that, as the Mott transition is approached, the Fermi-liquid coherence temperature continuously falls to the scale of kelvins with divergent quasi-particle decay rate in the metal side and the charge gap gradually closes in the insulator side. The Clausius-Clapeyron analysis of the pressure-temperature phase diagram provides thermodynamic evidence for the extremely weak first-order nature of the Mott transition. These results suggest that the spin liquid hosts a spinon Fermi surface, which turns into an electron Fermi surface when charges are Mott delocalized.**

In a Mott insulator, mutually interacting spins are generally ordered at low temperatures. However, antiferromagnetic interactions are self-conflicting for spins on a triangle-based lattice, which may fall into exotic states called quantum spin liquids[5,6]. Indeed, spin liquids were substantiated in the organic materials with triangular lattices, $\kappa$-(ET)$_2$Cu$_2$(CN)$_3$ [ET denotes BEDT-TTF = bis(ethylenedithio)tetrathiafulvalene] and EtMe$_3$Sb[Pd(dmit)$_2$]$_2$ (Refs. 7 and 8); they exhibit unconventional spin excitations of gapless or marginally gaped nature[7–13]. These properties lead to an intriguing proposal of delocalized spin excitations—spinons with a Fermi surface[14,15], which has recently been discussed in a strong spin-orbit-coupled triangular-lattice system YbMgGaO$_4$ as well[16,17].

In both organic materials, the spin liquids reside near the Mott metal-insulator transition, which is accessible by applying pressure[18–21]. The vicinity to the Mott transition is theoretically suggested to be essential to the emergence of spin liquids in triangular lattices[6,14]. In a sense that spin ordering does not follow, the metal-insulator transition of spin liquids is a *genuine* Mott transition, at which only the charge degrees of freedom are frozen, as Mott originally conceptualized. If the spin excitations in the spin liquid is delocalized as suggested theoretically, the itinerancy of the electrons is lost in the charge sector on the Mott transition but persists in the spin sector as *fractionalized spinons* that hold a Fermi surface, leading to unconventional Mott localization of electrons; indeed, a continuous Mott transition distinct from the conventional cases is predicted[3,4,22]. Thus, the close examination of the Mott transition is expected to provide new insight into fractionalization of particles in a condensed matter. Here, we investigate the charge transport of κ-(ET)$_2$Cu$_2$(CN)$_3$ under precisely controlled pressure to reveal how the Fermi liquid and the spin liquid meet in the pressure ($P$)−temperature ($T$) plane.

κ-(ET)$_2$Cu$_2$(CN)$_3$ is a layered organic system comprising of conducting ET layers and insulating Cu$_2$(CN)$_3$ layers. In the conducting layers, the ET molecules form dimers, which constitute a nearly isotropic triangular lattice with a half-filled band[23] (Fig. S3). The in-plane resistance was measured with the standard dc four-probe method. Hydrostatic pressure was applied by using, as pressure medium, helium gas for $P < 200$ MPa and Daphne 7373 oil or DEMNUM S-20 oil for $P > 200$ MPa. Experimental details are described in Methods.

The main points of the present results for κ-(ET)$_2$Cu$_2$(CN)$_3$ are featured in the $P$-$T$ phase diagram (Figs. 1a-c), which was constructed on the basis of the resistivity data described below (Figs. 2a-d). The first-order transition curve separating the spin-liquid insulator and a metal (or a superconductor) shows a characteristic bending profile (Fig. 1a) and leads to a fan-shaped quantum critical region at high temperatures[19] (Figs. 1b,c). The temperature region of a Fermi liquid, in which the resistivity follows the $T^2$-law, is strongly suppressed as the Mott boundary is approached. The proximity of the superconducting phase to the spin-liquid phase is confirmed. Figure 2a shows the temperature dependence of resistance for $P = 0$–202 MPa, where a systematic variation from an insulating behavior to a metallic behavior emerges. For $P \geq 135$ MPa, the resistance exhibits jumps with barely recognizable hysteresis, indicating a weak first-order Mott transition; we define $P_c = 135$ MPa as the critical pressure of the Mott transition at low temperatures. Absence of resistance jump for $T \geq 20$ K (Fig. 2d) means that the critical endpoint of the Mott transition is located in between 16 K and 20 K, which is about a half of the critical endpoint in κ-(ET)$_2$Cu[N(CN)$_2$]Cl, 40 K (Ref. 24).

Figure 2b shows the resistance in the metallic phase, plotted against $T^2$. The $T^2$ dependence expected in a conventional Fermi liquid, $R(T) = R_0 + AT^2$, holds in a limited low-temperature range below the coherence temperature $T^*$, which is pushed down to the order of kelvin near the Mott boundary (Figs. 1a-c). The three orders of magnitude drop of $T^*$ from the bare energy scale of the bandwidth, ~0.5 eV (~6000 K)[20], suggests that the system is situated near the quantum critical regime. Referring to theoretical investigations[4,22], $T^*$ might correspond to the crossover temperature between Fermi liquid and marginal Fermi liquid, in which the quasi particles are scattered by critical fluctuations of gauge field. The locations of $T^*$ and the quantum critical region in the pressure-temperature phase diagram (Fig. 1c) are overall consistent with the theoretical prediction. The coefficient $A$ of the $T^2$ term, normalized to the room-temperature resistance value of each sample, is enhanced as the Mott boundary is approached (Fig. 3), indicating an increase in quasi-particle decay rate. The pressure dependence of $A$ is well fitted by a power-law, $A \propto |P-P_c|^{-0.75}$, which is compared with the theoretical prediction[4] of the quasi-particle decay rate near the quantum Mott transition, $\gamma_{qp} \propto |P-P_c|^{-2\nu}$ ($2\nu \sim 1.34$); the agreement of the exponents is not excellent. The gentler divergence in the experiment than the theory may reflect the weak first-order nature of the transition. In the conventional correlated-electron systems, $A$ is proportional to the square of effective mass and thus a power-law divergence of $A$ is linked to a power-law divergence of the effective mass. Note, however, that this may not be the case in the quantum Mott transition of a spin liquid, according to quantum-field[4] and phenomenological[25] theories; namely, the effective mass may not diverge in a power-law manner.

In the insulating phase, the resistance is characterized by a charge gap, $\Delta$, obtained by fitting the form of $R(T) = C \exp(\Delta/2T)$ to the data (Fig. 2c). A remarkable feature is the gradual closing of the charge gap following $\Delta \propto |P-P_c|^{0.71}$ for $|P-P_c| \gtrsim 25$ MPa, which suggests the quasi-continuous nature of the Mott transition (Fig. 3). This form is close to the theoretical prediction, $\Delta \propto |P-P_c|^{0.67}$, for the quantum Mott transition[4,22], while some numerical studies suggest $\Delta \propto |P-P_c|$ (Refs. 26,27). The anomalous drop of $\Delta$ for $|P-P_c| < 25$ MPa implies marginally gapped or gapless charge excitations in the

vicinity of the Mott transition; indeed, the resistance shows a plateau in the narrow range of pressure around the Mott boundary below 10 K (Figs. 2a,d). To visualize these resistance behaviors, the logarithm of resistance is indicated by a range of color on the *P-T* phase diagram (Fig. 2e), which shows that the (yellow-colored) plateau region resides in the lower-pressure side of the Mott boundary. Remarkably, such a resistance plateau is theoretically suggested to appear at the two-dimensional quantum Mott transition of a spin liquid[4,22]. The predicted sheet resistance at the plateau is ~ 8 $h/e^2$ ( $h$ is the Plank constant and $e$ is the elementary charge), which is smaller than the experimental sheet resistance of one conducting layer in the plateau, ~33 $h/e^2$ (Supplementary Information). In the experiments, the plateau appears in a narrow but finite pressure region while it occurs only at a critical pressure, although fan-shaped at high temperatures, in the theoretical prediction. The disorder in real systems can extend the critical point into a finite region, which may also explain the discrepancy in the magnitude of the sheet resistance. It is suggested that disorder in the polymeric atomic configurations in the anion layers affects the charge transport at ambient pressure[28]

As seen above, the transport characteristics around the Mott boundary are featured by a substantial falloff in *T**, the critical behaviors of the coefficient *A*, the closing of charge gap *Δ*, and the resistance plateau, all of which go against the conventional first-order Mott transition but accord with the theoretical consequences of the Mott transition of a spin liquid with a spinon Fermi surface[3,4,22]. Below, we discuss thermodynamics of the Mott phase diagram in terms of the Clausius-Clapeyron relation, which relates the slope of the phase boundary, $\partial P_{jump}/\partial T_{jump}$, to the entropy difference between the neighboring phases through $S_{QSL} - S_{metal} = \Delta V \times \partial P_{jump}/\partial T_{jump}$, where $S_{QSL}$ ($S_{metal}$) is the molar entropy of the spin liquid (metal) and $\Delta V$ (= $V_{QSL} - V_{metal}$) is the molar volume change. Assuming that the lattice contributions to $S_{QSL}$ and $S_{metal}$ do not largely differ, $S_{QSL} - S_{metal}$ is approximately equal to $S_{QSL,spin} - S_{metal,el}$, where $S_{QSL,spin}$ ($S_{metal,el}$) is the spin (electron) contribution to $S_{QSL}$ ($S_{metal}$); the validity of this assumption is argued later. The $\partial P_{jump}/\partial T_{jump}$ determined from the fitting curve of the phase boundary (Fig. 1a) gives the temperature dependence of $S_{QSL,spin} - S_{metal,el}$ with $\Delta V$ as a parameter (Fig. 4a). Because $\Delta V$ and $P$ are thermodynamically conjugate variables, $\Delta V$ plays a role of order parameter of Mott transition and therefore measures the strength of the first-order transition.

For evaluating $S_{QSL,spin}$, $S_{metal,el}$ is reasonably assumed to be $\gamma T$, where $\gamma$ is estimated at ~ 27.6 mJ K$^{-2}$ mol$^{-1}$, referring to the spin susceptibility of $\chi_{metal}$ = 3.8 × 10$^{-4}$ emu mol$^{-1}$ (Ref. 29) and the Wilson ratio of $R_W$ = 1, which holds in the metallic phases in κ-type compounds (Supplementary Information). The entropy of the spin liquid is obtained through $S_{QSL,spin} = \gamma T + \Delta V \times (\partial P_{jump}/\partial T_{jump})$, as displayed for several values of $\Delta V$ in Fig. 4b. The $S_{QSL,spin}$ decreases on cooling particularly below 10 K and falls below $S_{metal,el}$ around 6 K (Fig. 4b). The rather steep entropy loss around 6 K implies that the 6-K anomaly observed at ambient pressure[9,10,30] persists up to the Mott boundary. Because $S_{QSL,spin}$ must not increase on cooling, $\Delta V$ should be ~ $2 \times 10^{-8}$ m$^3$ mol$^{-1}$ or less (Fig. 4b).

For comparison, we performed a similar analysis for κ-(ET)$_2$Cu[N(CN)$_2$]Cl which exhibits an antiferromagnetic order in the Mott insulating phase (Figs. 4c,d and Fig. S5). The spin entropy of the antiferromagnetic insulator (AFI) side, $S_{AFI,spin}$, rapidly decreases with temperature, even falling below $S_{metal,el}$ at as high as 30 K, irrespectively of the choice of the $\Delta V$ value. In the insulating side on the Mott boundary, spins undergo an antiferromagnetic order at approximately $T_N = 15$ K (Ref. 31). Thus, spins lose their entropy from much higher temperatures than the Néel order due to short-range ordering, as shown by the thermodynamic and magnetic measurements at ambient pressure[32,33]. The requirement that the entropy should die out rapidly at low temperatures narrow down the possible value of $\Delta V$ to a range of 4–6 $\times 10^{-7}$ m$^3$ mol$^{-1}$ (Fig. 4d), which is in good agreement with the experimentally determined value of ~ $5.4 \times 10^{-7}$ m$^3$ mol$^{-1}$ (Ref. 34, see also Supplementary Information) and thus corroborates the validity of the present analysis. Remarkably, the $\Delta V$ of κ-(ET)$_2$Cu$_2$(CN)$_3$ is several ten times smaller than the $\Delta V$ of κ-(ET)$_2$Cu[N(CN)$_2$]Cl, which provides thermodynamic evidence for the extremely weak first-order nature of the Mott transition in κ-(ET)$_2$Cu$_2$(CN)$_3$, in accord with the quasi-continuous nature of the transition revealed by the charge transport. Note that the large value of $\Delta V$ in κ-(ET)$_2$Cu[N(CN)$_2$]Cl holds even above $T_N$, meaning that the extremely small $\Delta V$ value in κ-(ET)$_2$Cu$_2$(CN)$_3$ is not attributable only to the paramagnetic nature of the Mott insulator.

The charge transport in κ-(ET)$_2$Cu$_2$(CN)$_3$ and the thermodynamic arguments consistently highlight the quasi-continuous nature of the Mott transition, which contradicts the conventional picture but is reconciled with the theoretical consequences of a transition from an electron Fermi surface to a spinon Fermi surface. We note, however, that the correspondence between the experiments and the theories is not exact in that κ-(ET)$_2$Cu$_2$(CN)$_3$ exhibits a first-order transition although it is extremely weak. This may indicate that some additional factors not included in the theoretical treatments affect the ground states; indeed, superconductivity appears in the metallic phase and the 6-K anomaly possibly suggestive of some instability of the spin liquid emerges in the insulating phase. Very recently, a DMFT-based theoretical treatment incorporating spinon excitations suggest the first-order nature of the Mott transition of a spinon Fermi surface. The faint remanence of the first-order nature may have some relevance to this theoretical consequence[35]. In any event, the present observations distinct from the conventional cases are expected to promote further theoretical studies.

**Methods:**

**Resistance measurements under pressure.**

Single crystals of κ-(ET)$_2$Cu$_2$(CN)$_3$ (named #1, #2, #3 and #4) were grown by the conventional electrochemical method. The in-plane resistivity was measured with the d.c. four-probe method. Gold wires of 15−25 μm in diameter were glued onto the crystal faces with carbon paste as electrodes. To apply hydrostatic pressure, we used the pressure mediums of helium (for samples #1 and #2) for $P < 200$ MPa, and Daphne 7373 oil (for sample #3) and DEMNUM S-20 oil (for sample #4) for $P > 200$ MPa. These pressure mediums are solidified at low temperatures. The melting temperature of helium is as low as 0−20 K for pressures studied here (Fig. S1), while those of the two oils are higher than 100 K. On the solidification of the pressure mediums, the pressure in the cell is somewhat reduced. The pressure reduction for helium medium, which is in a range of 10−30 MPa, could be precisely evaluated, using the method described in Supplementary Information. The sample #1 was studied across the helium solidification curve in Fig. S1; the pressure was determined by the method with an accuracy of 1 MPa. Sample #2 was measured at higher temperatures than the helium solidification curve. As for samples #3 and #4 put in the oil mediums, the pressure values quoted in this Letter are those obtained by subtracting 300 MPa (sample #3) or 250 MPa (sample #4) from the pressure monitored at room temperature.

**Figures**

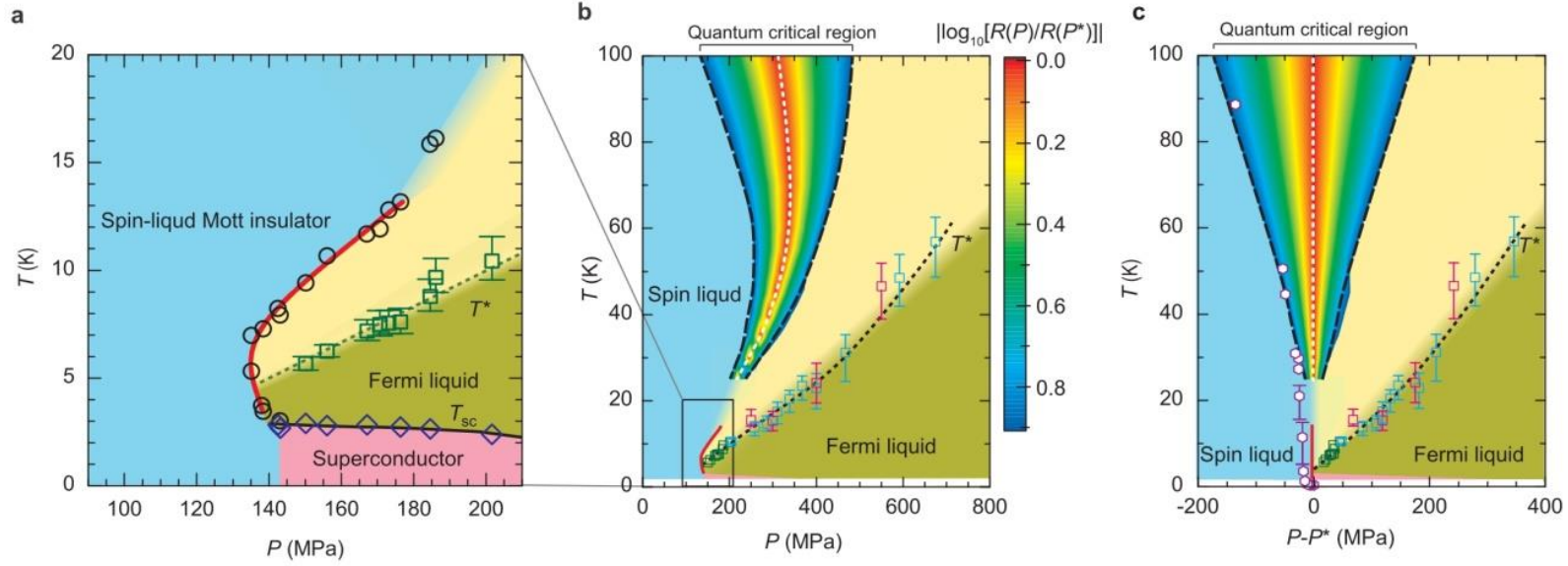

**Figure 1 | Pressure-temperature phase diagram of κ-(ET)$_2$Cu$_2$(CN)$_3$. a,** Low-temperature phase diagram in the vicinity of the Mott transition constructed from the resistance behavior of samples #1 and #2. The black open circles indicate the points, at which resistive jumps associated with the first-order Mott transition are observed, and the red bold line is a fit to the data points, which is used in the entropy analyses (Figs. 4a-d). The blue diamonds indicate the onset of the superconducting transition $T_{sc}$. The green squares indicate the temperatures $T^*$, at which the resistance deviates from the $T^2$ law characteristic of the Fermi liquids. **b,** Phase diagram in a wide temperature-pressure range. The squares indicate the $T^*$ values determined from the resistance behavior of sample #1(green), #3(pink), and #4(blue). The white broken line indicates the Mott crossover pressure (the Widom line), $P^*(T)$, taken from Ref. 19. The fan-shaped contour plot, which is reproduced from Ref. 19, represents the magnitude of $|\log_{10} R(P)/R(P^*)|$ in the quantum critical region. **c,** Phase diagram in $P$-$P^*$ – $T$ plane. The open hexagons indicate the charge gaps (Fig. 3) divided by a factor three, $\Delta/3$, in order to see the correspondence with the pressure profile of the quantum-critical energy scale. It is observed in the figure that all of the energy scales characteristic of the metallic and insulating phases go to collapse toward the Mott transition point.

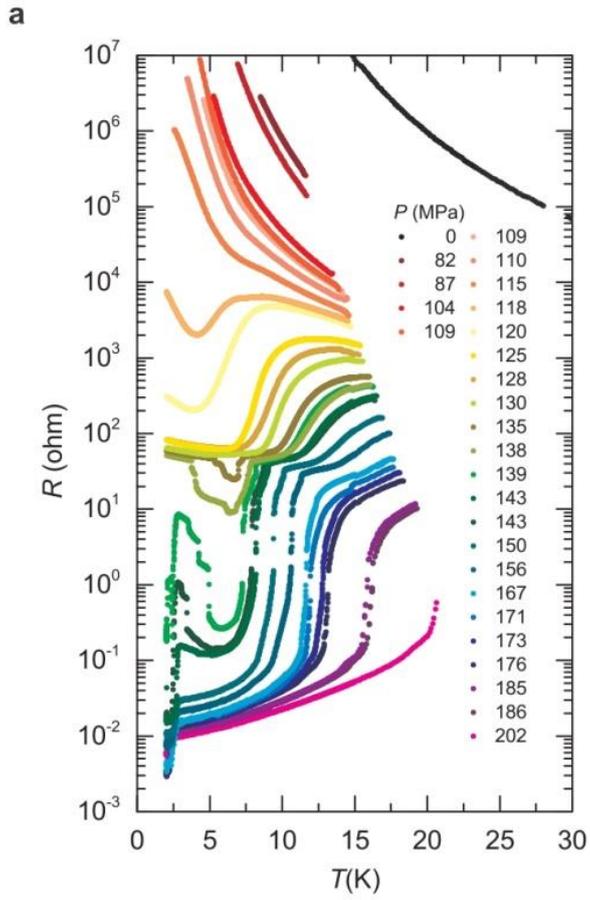
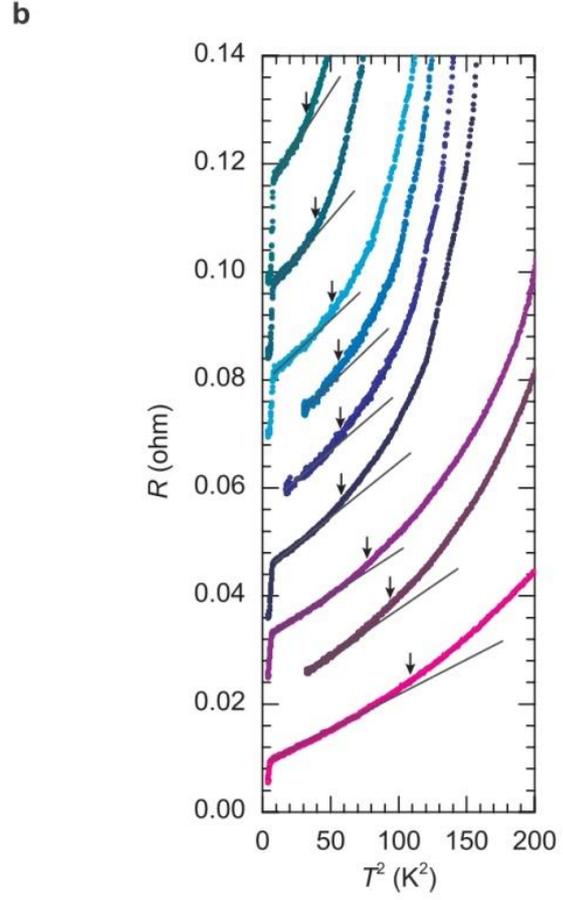
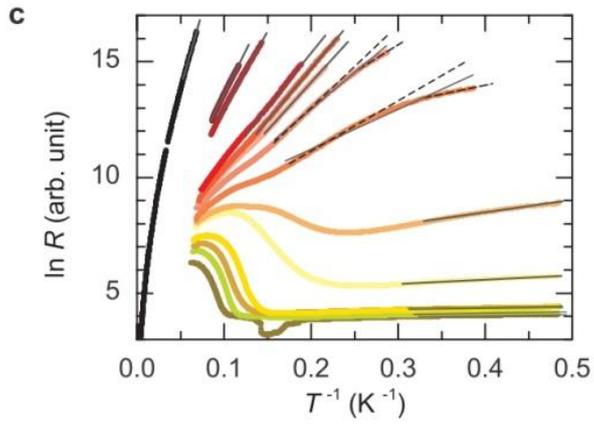
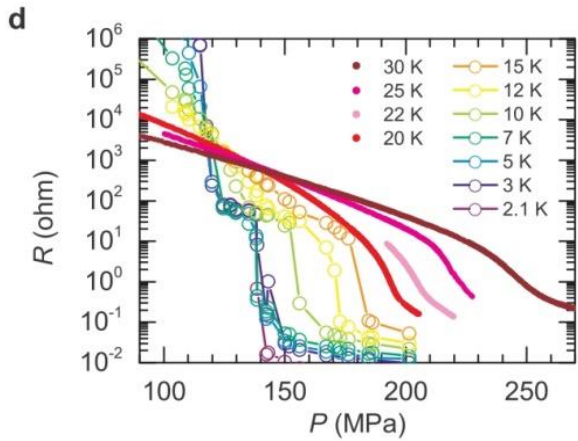
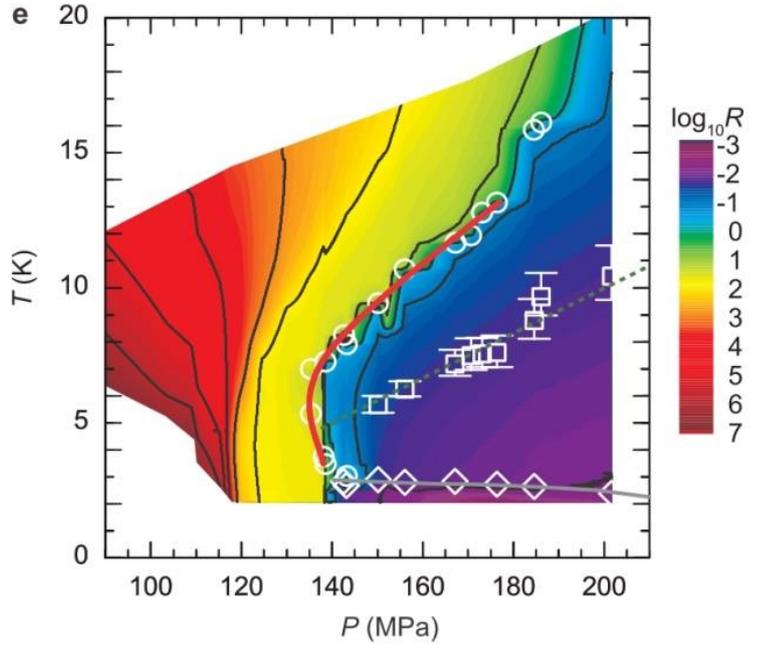

**Figure 2 | Transport properties of κ-(ET)$_2$Cu$_2$(CN)$_3$ near the Mott transition. a,** Temperature dependence of resistance of sample #1. The presented data are for low temperatures below the melting point of the helium pressure medium because the solidification is followed by a pressure drop, causing a spurious change in resistivity (see Supplementary Information). **b**, Resistance (offsetted by an interval of 0.01 ohm) vs. $T^2$. The solid lines are the fits of the form, $R(T) = R_0 + AT^2$, to the data. The crossover temperature $T^*$ at each pressure is indicated by arrows (see Supplementary Information for the definition). **c,** Arrhenius plot of resistance in the insulating phase. The solid lines are the fits of the form, $\ln R(T) = \Delta/2 \times T^{-1} +$ const. , to the data at low temperatures. As for $P = 110$ MPa and 115 MPa, where data are not on straight lines, we additionally performed the fitting for the high (low) temperature part of the data indicated by broken lines to evaluate the higher (lower) bound of the errors in $\Delta$. **d,** Pressure dependence of the resistance of sample #1 ($T \leq 22$ K) and #2 ($T \geq 25$ K) at fixed temperatures. The resistance of sample #2 was multiplied by 1.5 for comparison. The measurements at $T$ above 20 K, where the helium pressure medium is in a liquid state, were performed in isothermal pressure sweeps. **e,** Contour plot of the logarithmic resistance of sample #1 in the $P$-$T$ plane. The black lines are contour lines. The open circles (Mott transition points), squares ($T^*$) and diamonds ($T_{sc}$) are reproduced from Fig. 1a.

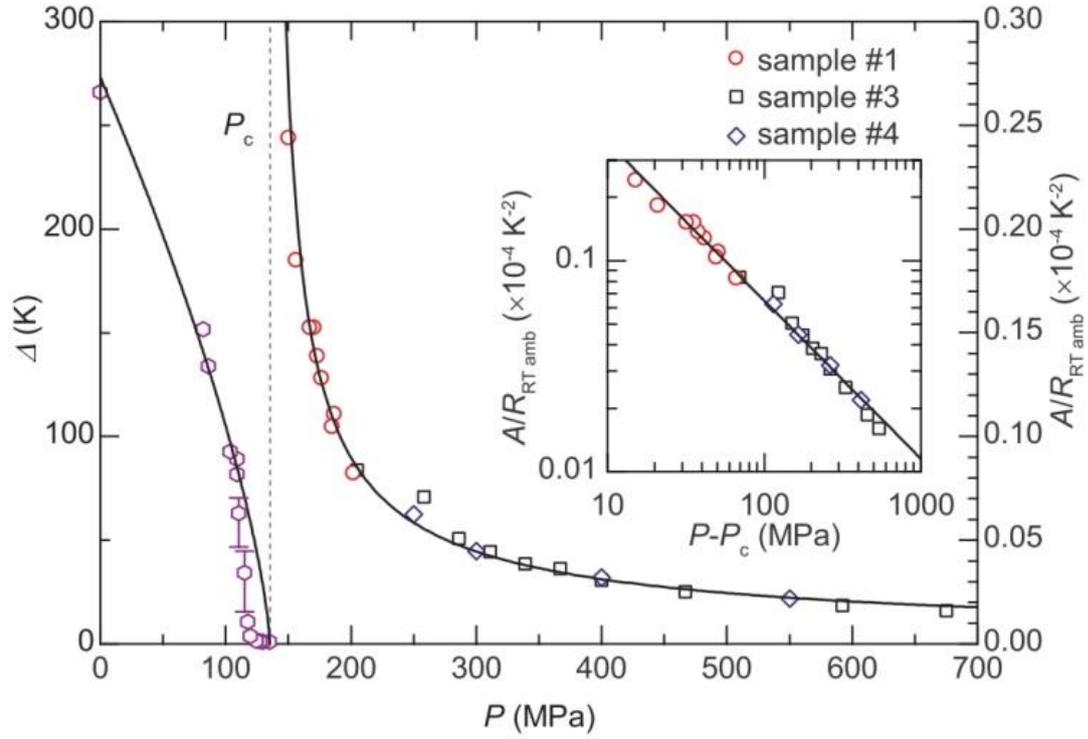

**Figure 3 | Pressure dependence of the charge gap $\Delta$ and the $T^2$ coefficient $A$.** The solid lines indicate the fitting curves of $\Delta \propto |P-P_c|^{0.71}$ and $A \propto |P-P_c|^{-0.75}$, where $P_c$ is 135 MPa. The $T^2$ coefficient $A$ is normalized by resistance at room temperature under ambient pressure, $R_{RT,ambient}$, for each samples. The inset shows the log-log plot of $A/R_{RT,ambient}$ vs. $P-P_c$.

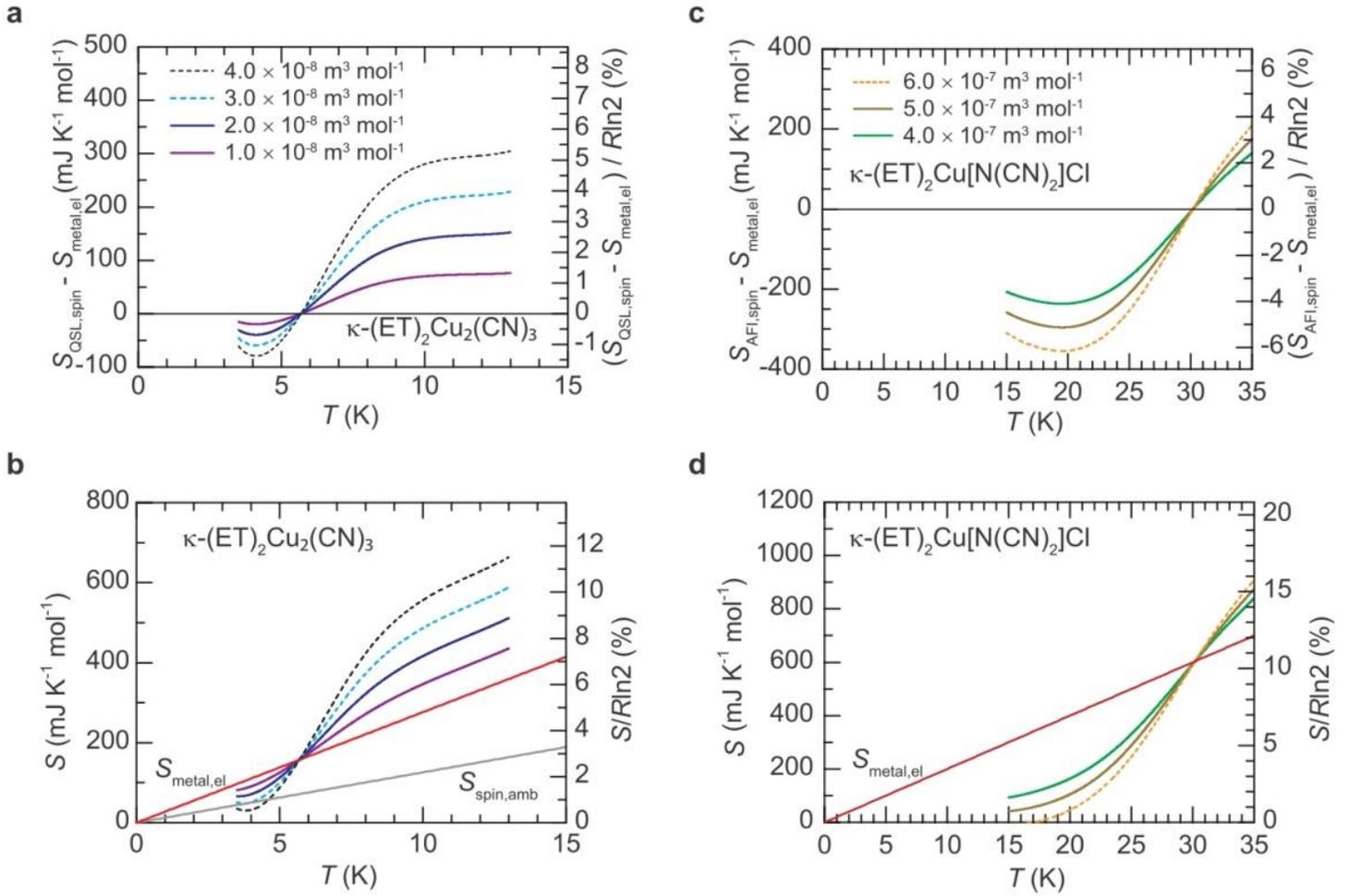

**Figure 4 | Thermodynamic properties of the spin liquid on the Mott boundary.** The temperature dependences of (**a**) $S_{QSL}$-$S_{metal}$ and (**b**) $S_{QSL,spin}$ for κ-(ET)$_2$Cu$_2$(CN)$_3$, and (**c**) $S_{AFI}$-$S_{metal}$ and (**d**) $S_{AFI,spin}$ for κ-(ET)$_2$Cu[N(CN)$_2$]Cl are indicated for several values of the parameter, $\Delta V$. $R$ is the universal gas constant. Dashed lines are unphysical results that do not fulfill the fundamental relation, $S > 0$ and $dS/dT > 0$, owing to too large $\Delta V$ values, which give the upper bounds for the realistic $\Delta V$ values. The gray line in (b) indicates $S_{spin,amb} = \gamma_s T$, with $\gamma_s = 12.6$ mJ K$^{-2}$ mol$^{-1}$ (Ref. 10). The red lines in (b) and (d) show $S_{metal,el} = \gamma T$, where $\gamma = 27.6$ mJ K$^{-2}$ mol$^{-1}$ for κ-(ET)$_2$Cu$_2$(CN)$_3$ and $\gamma = 20$ mJ K$^{-2}$ mol$^{-1}$ for κ-(ET)$_2$Cu[N(CN)$_2$]Cl. The $\gamma$ value in the latter is that of the partially deuterated κ-(ET)$_2$Cu[N(CN)$_2$]Br (Ref. 36), which is a Mott insulator on the verge of the Mott transition.

**Acknowledgments:**

This work was supported in part by JSPS KAKENHI under Grant Nos. 20110002, 25220709, 24654101, 25287080, and 17K05532 and the US National Science Foundation under Grant No. PHYS-1066293 and the hospitality of the Aspen Center for Physics.

**Author Contribution:**

T.F. and K.Kanoda designed the experiments. T.F., K.Kobashi and Y.K. performed the resistance measurements. K.M. grew the single crystal for the study. All authors interpreted the data. T.F. wrote the manuscript with the assistance of K.M. and K.Kanoda. K.Kanoda headed this project.

**Additional information**

Correspondence and requests for materials should be addressed to T.F. or K.Kanoda.

**Supplementary information**

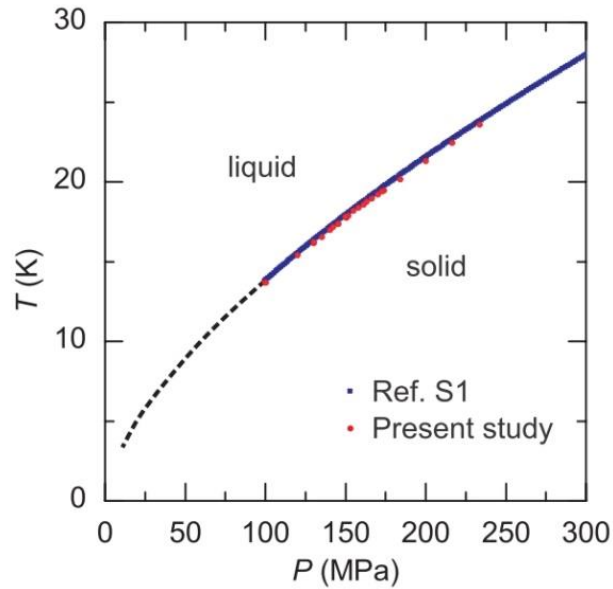

**Figure S1 | Phase diagram of helium.**

**Determination of internal pressure below the helium solidification temperature**

The *P-T* phase diagram of helium is displayed in Fig. S1. Using helium as a pressure medium, we varied pressure continuously in the *P-T* region above the solidification line. Below this line, the pressure is inevitably fixed and only temperature variation is available in experiments. As described in Methods, the internal pressure is reduced by approximately 10–30 MP on the solidification of helium in the present pressure apparatus. There was a small difference between the internal pressure in the solid helium and the indication of pressure gauge, which measures the pressure of the helium gas linked to solid helium through a tube (Note that there is no difference between them when helium is in a liquid or gas state). Nevertheless, we could precisely estimate the internal pressure by monitoring the resistance of the sample by the following method. Figure S2 shows examples of the resistance behaviors (of sample #1) on cooling across the helium solidification temperature. The helium starts to solidify at a temperature indicated by an arrow labelled $R_{solid}$ and the solidification is completed at a temperature pointed by $R_{solid}$'. In between the two temperatures, the internal pressure is gradually reduced along the solidification curve and the resistance of the sample increases; at lower temperatures, the internal pressure is fixed. We can know the internal pressure at the temperature marked by $R_{solid}$' and lower by the following procedure. First, we measured the resistance on the solidification curve, $R_{solid}(P)$, by reading the pressure and resistance values just before the solidification, $P_{solid}$ and $R_{solid}$, which should be equal to the internal pressure and the resistance at the indicated pressure, respectively, on the solidification line (Fig. S2). Second, we fitted the $P_{solid}$ vs. $R_{solid}$ data set by fifth-order polynomial, which gives the form of $R_{solid}(P)$ on the solidification curve. Finally, by using the fitting curve of $R_{solid}(P)$, we know the internal pressure, $P_{solid}$', which gives the resistance just after the completion of the helium solidification, $R'_{solid}$.

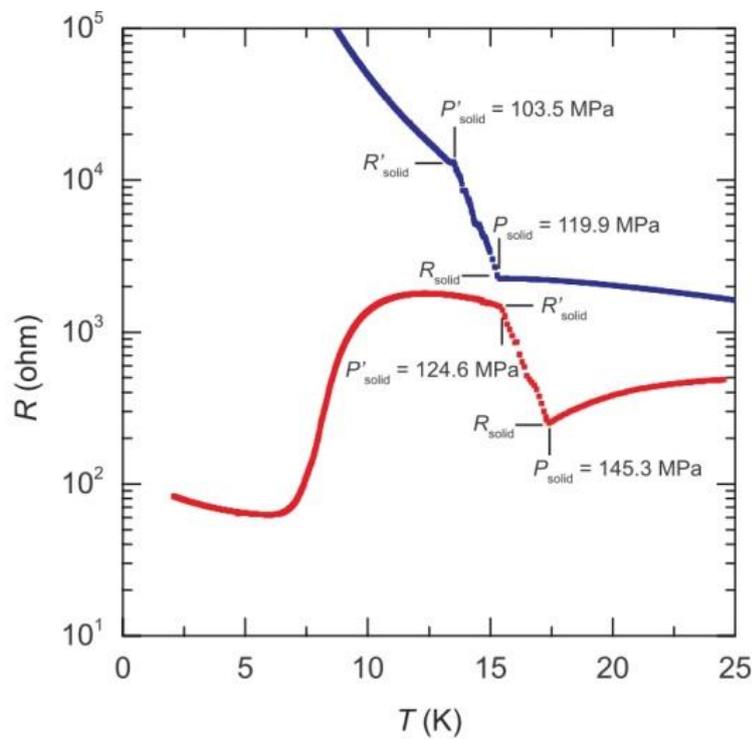

**Figure S2 | Examples of temperature dependence of resistance across the solidification temperature of helium.**

**Crystal structure of κ-(ET)$_2$Cu$_2$(CN)$_3$**

Figure S3 shows the structure of conducting layer in κ-(ET)$_2$Cu$_2$(CN)$_3$. The conducting layers comprised of (ET)$_2$ dimers are stacked alternately with the non-magnetic insulating Cu$_2$(CN)$_3$ layers. A (ET)$_2$ dimer plays a role of one lattice point of an anisotropic triangular lattice and accommodates one hole. Due to the strong on-site Coulomb repulsion, this material is a Mott insulator at ambient pressure.

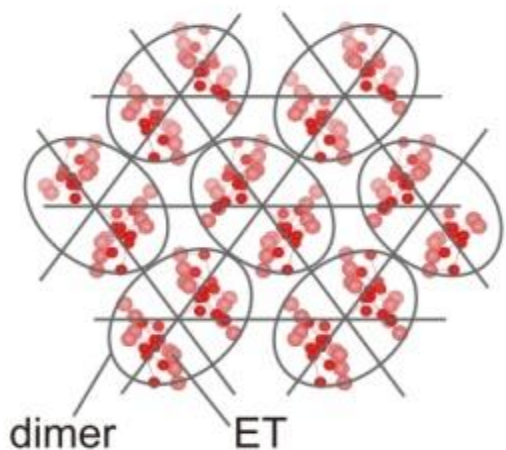

**Figure S3 | Structure of conducting layer in κ-(ET)$_2$Cu$_2$(CN)$_3$.**

**Determination of the crossover temperature $T^*$**

We defined $T^*$ as the temperature at which the deviation of resistance $R(T)$ from $R_{\text{fit}}(T) = R_0 + AT^2$ exceeds 10% of $AT^2$, which fits the low-temperature data. We evaluated the upper and lower bounds of errors in $T^*$ (Fig. 1) as the temperatures giving the 15% and 5% deviations of $R(T)$, respectively. Figure S4 displays the resistance of samples #3 and #4 against $T^2$.

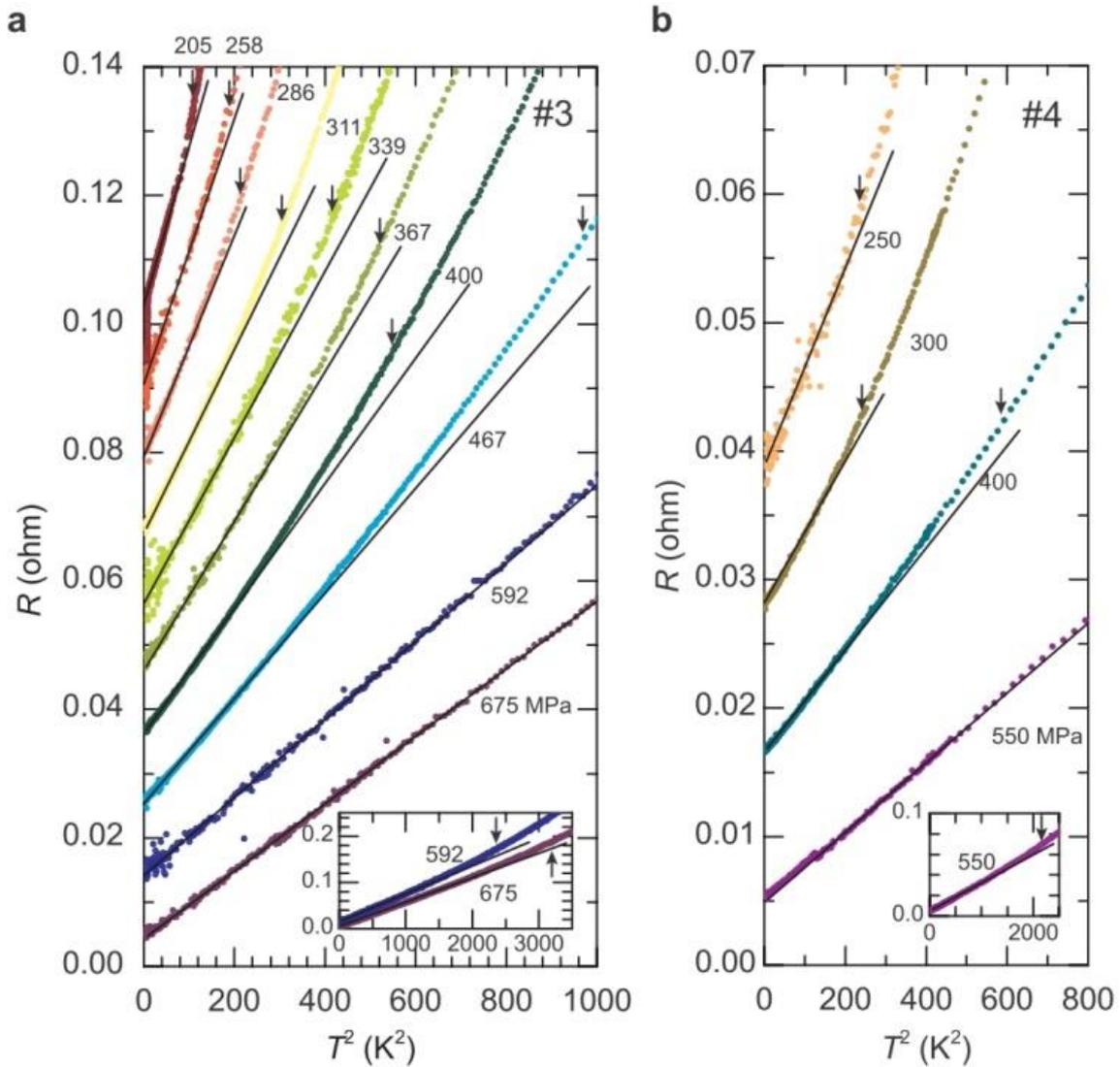

**Figure S4 | Resistance vs. $T^2$ for sample #3 (a) and sample #4 (b).** The resistance is plotted with a vertical offset of 0.01 ohm interval for clarity The solid lines are the fits of the equation of $R(T) = R_0 + AT^2$ to the low-temperature parts of the data. Arrows indicate $T^{*2}$ at each pressure.

**Wilson ratio of the metallic κ-(ET)$_2$X situated on the Mott boundary**

We use the value of the Wilson ratio $R_W$ of the κ-(ET)$_2$X metallic phase neighbouring the Mott insulating phase in the analysis for deducing the spin entropy in κ-(ET)$_2$Cu$_2$(CN)$_3$. $R_W$ is given by $R_W = (\pi^2 k_B^2/3\mu_B^2) \times (\chi/\gamma)$, where $k_B$ is the Boltzmann constant, $\mu_B$ is the Bohr magneton, $\chi$ is spin susceptibility and $\gamma$ is electronic specific heat coefficient. The experimental data of $\chi$ and $\gamma$ are available for two compounds with $X$ = Cu(NCS)$_2$ and Cu[N(CN)$_2$]Br. For $X$ = Cu(NCS)$_2$, $\gamma$ of ~ 24 mJ K$^{-2}$ mol$^{-1}$ (Ref. S2) and $\chi_{metal}$ of ~ 3.2 × 10$^{-4}$ emu mol$^{-1}$ (Ref. S3) yields $R_W$ of ~ 0.97. For $X$ = Cu[N(CN)$_2$]Br, $\gamma$ of ~ 22 mJ K$^{-2}$ mol$^{-1}$ (Ref. S4) and $\chi_{metal}$ of ~ 3.0 × 10$^{-4}$ emu mol$^{-1}$ (Ref. S5) yields $R_W$ of ~ 0.99. Thus, $R_W$ of the κ-(ET)$_2$X metallic phase is assumed to be approximately unity.

**Volume change $\Delta V$ on the Mott transition in $\kappa$-(ET)$_2$Cu[N(CN)$_2$]Cl**

Figure S5(a) shows the temperature-pressure phase diagram of $\kappa$-(ET)$_2$Cu[N(CN)$_2$]Cl reproduced from Ref. S6. Figure S5(b) shows a fitting curve of the first-order Mott transition points in Fig. 5(a). In the main text, we performed the Clausius-Clapeyron analysis for $\kappa$-(ET)$_2$Cu[N(CN)$_2$]Cl by using the slope of the transition curve in Fig. S5(b).

As described in the main text, the Clausius-Clapeyron analysis for $\kappa$-(ET)$_2$Cu[N(CN)$_2$]Cl give an evaluation of the molar volume change at the first-order Mott transition, $\Delta V$, which yields $4-6 \times 10^{-7}$ m$^3$ mol$^{-1}$. On the other hand, $\Delta V$ in $\kappa$-(ET)$_2$Cu[N(CN)$_2$]Cl can be independently evaluated by using the expansivity data reported in Ref. 34. The volume change on the Mott transition is approximately expressed by $\Delta V/V \sim \Delta l_a/l_a + \Delta l_b/l_b$, where $l_{i=a,b}$ and $\Delta l_{i=a,b}$ are the molar length of a sample and the change of it on the Mott transition measured along the in-plane $a$ axis and the out-of-plane $b$ axis. According to Ref. 34, $\Delta l_c/l_c$ is negligible, compared with $\Delta l_a/l_a$ and $\Delta l_b/l_b$; namely, $\Delta l_c/l_c \ll \Delta l_a/l_a, \Delta l_b/l_b$. Referring to Fig. 2 in Ref. 34, $\Delta l_a/l_a \sim 3.2 \times 10^{-4}$ and $\Delta l_b/l_b \sim 2.3 \times 10^{-4}$ on the Mott-transition at 22 MPa and 30 K. Then, $\Delta V$ of $\kappa$-(ET)$_2$Cu[N(CN)$_2$]Cl at 30 K is estimated to be $5.4 \times 10^{-7}$ cm$^3$ mol$^{-1}$. It is considered that $\Delta V$ nearly saturates at 30 K, which is well below the critical end-point, $T_c = 40$ K.

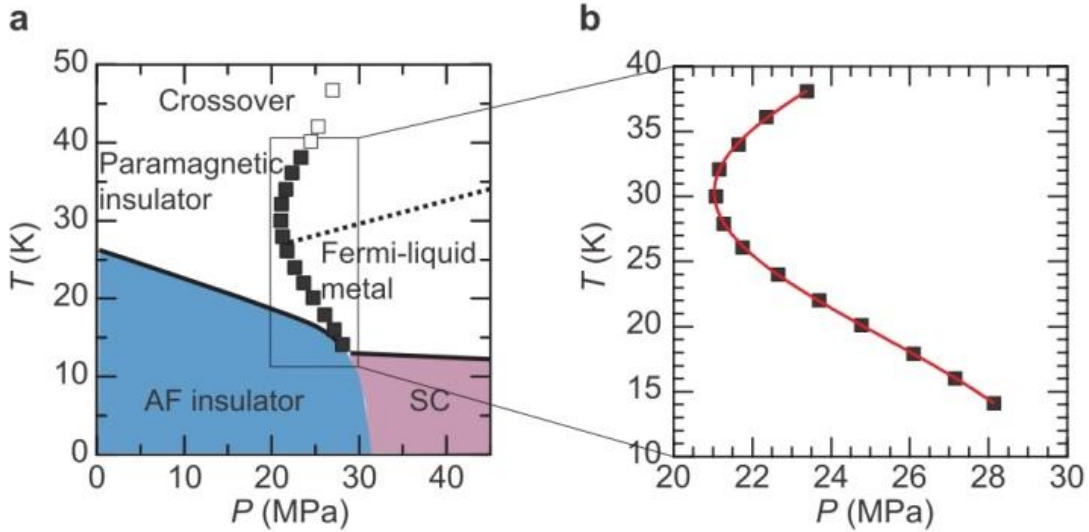

**Figure S5 | Pressure-temperature phase diagram of $\kappa$-(ET)$_2$Cu[N(CN)$_2$]Cl. a,** Phase diagram near the Mott boundary. Close and open squares indicate the pressures at the first-order Mott transition and the Mott crossover, respectively. The data points were reproduced from Ref. S6. **b,** The first-order transition curve that fits to the data points

reproduced from Ref. S6.

**Sheet resistance near the Mott boundary**

As $\kappa$-(ET)$_2$Cu$_2$(CN)$_3$ is structured by the stacking of two-dimensional layers of $d = 1.45$ nm in thickness, the sheet resistance of the individual layer, $R_{\text{sheet}}$, at the resistance plateau is given by $R_{\text{sheet}} = \rho_{\text{plateau}} \times d^{-1} = R_{\text{plateau}} \times \rho_{\text{amb,RT}} / R_{\text{amb,RT}} \times d^{-1}$, where $\rho_{\text{plateau}}$ is the resistivity at the plateau and $\rho_{\text{amb,RT}}$ ($R_{\text{amb,RT}}$) is resistivity (rersistance) at room temperature under ambient pressure. Using the values of $R_{\text{plateau}} = 60\ \Omega$, $R_{\text{amb,RT}} = 14.6\ \Omega$ and $\rho_{\text{amb,RT}} = 30\ \text{m}\Omega\text{cm}$ (Ref. 19), we obtain $R_{\text{sheet}} = 8.2 \times 10^5\ \Omega = 33 \times h/e^2$.

**Supplementry References**